\documentclass[camera]{jpaper}

\usepackage{gensymb}
\usepackage{subfig}
\usepackage{multirow}
\usepackage{mathtools}
\usepackage{xspace}
\usepackage{textcomp}
\usepackage{listings}
\usepackage[plain]{algorithm}  % the floater wrapper for algorithmic
\usepackage[noend]{algpseudocode}
\usepackage{algorithmicx}
\floatname{algorithm}{Template}
\usepackage{xfrac}
\usepackage{subscript}

\usepackage{dblfloatfix}
\usepackage{appendix}

\usepackage{float}
\usepackage{booktabs}
\usepackage{placeins}
\usepackage{balance}
\usepackage[absolute]{textpos}
\usepackage[nocompress]{cite}
\usepackage{array}

\makeatletter
\let\MYcaption\@makecaption
\makeatother

\makeatletter
\let\@makecaption\MYcaption
\makeatother

\usepackage{fixltx2e}
\usepackage{dblfloatfix}
\usepackage[nolessnomore, italic]{mathastext}
\usepackage[T1]{fontenc}
\usepackage[usenames,dvipsnames,svgnames,table]{xcolor}
\usepackage[normalem]{ulem}
\usepackage{enumitem}
\usepackage{setspace}
\usepackage{indentfirst}
\usepackage{footmisc}
\usepackage{fancyhdr}
\usepackage{authblk}
\usepackage[us,12hr]{datetime}
\usepackage[keeplastbox]{flushend}
\usepackage[hidelinks]{hyperref}

\newif\ifcameraready
\camerareadytrue
\newcommand{\versionnum}[0]{5}

\ifcameraready
  \newcommand{\new}[0]{}
  \newcommand{\newI}[0]{}
\else
  \newcommand{\new}[1]{{\color{blue}#1}}
  \newcommand{\newI}[1]{\textcolor{BrickRed}{#1}}
\fi

\usepackage{ifthen}
\usepackage{calc}
\usepackage{pifont}
\usepackage{color}
\usepackage{fancyhdr}
\usepackage{tikz}
\usepackage{anyfontsize}

\newcounter{hours}
\newcounter{minutes}

\newboolean{timeofmake}
\setboolean{timeofmake}{false}

\newcommand{\ignore}[1]{}

\newcommand{\villa}[0]{{VILLA-DRAM}\xspace}

\newcommand{\acro}[0]{{LISA}\xspace}
\newcommand{\lisa}[0]{{LISA}\xspace}
\newcommand{\lisarc}[0]{{LISA-RISC}\xspace}

\newcommand{\lisapre}[0]{{LISA-LIP}\xspace}

\newcommand{\lisavilla}[0]{{LISA-VILLA}\xspace}

\newcommand{\lisarcvilla}[0]{{LISA-(RISC+VILLA)}\xspace}
\newcommand{\baseline}[0]{\texttt{memcpy}\xspace}

\newcommand{\xferfull}[0]{{row buffer movement}\xspace}

\newcommand{\xfer}[0]{{RBM}\xspace}

\newcommand{\rc}[0]{{RowClone}\xspace}

\newcommand{\capfullname}[0]{{Low-Cost Inter-Linked Subarrays}\xspace}
\newcommand{\fullname}[0]{{low-cost inter-linked subarrays}\xspace}

\newcommand{\lisafullname}[0]{{\underline{L}ow-cost \underline{I}nter-linked
\underline{S}ub\underline{A}rrays}\xspace}

\newcommand{\rcirsa}[0]{{RC-InterSA}\xspace}

\newcommand{\figputGHWHeight}[3]{
\begin{figure}[h]
\begin{minipage}{\linewidth}
\footnotesize %also for forcing a baselinestretch update
\begin{center}
\includegraphics[width=1.0\linewidth,height=#3in]{plots/gnuplots/#1}%
\end{center}
\vspace{-0.15in}
\caption{#2 \label{fig:#1}}
\end{minipage}
\end{figure}
}

\newcommand{\figputGTW}[2]{
\begin{figure}[!t]
\begin{minipage}{\linewidth}
\begin{center}
\includegraphics[width=1.0\linewidth]{plots/gnuplots/#1}
\end{center}
\vspace{-0.15in}
\caption{#2 \label{fig:#1}}
\end{minipage}
\end{figure}
}

\newcommand{\figref}[1]{Figure~\ref{fig:#1}}
\newcommand{\tabref}[1]{Table~\ref{tab:#1}}

\newcommand{\ssecref}[1]{Section~\ref{ssec:#1}}

\title{LISA: Increasing Internal Connectivity in DRAM\\ for Fast Data Movement
  and Low Latency}

\author{%
{Kevin K. Chang$^{1,2}$}%
\qquad%
{Prashant J. Nair$^{3,4}$}%
\qquad%
{Saugata Ghose$^{2}$}%
\vspace{2pt}\\
{Donghyuk Lee$^{5,2}$}%
\qquad%
{Moinuddin K. Qureshi$^{4}$}%
\qquad%
{Onur Mutlu$^{6,2}$}}%
\affil{%
{\it%
$^{1}$Facebook\qquad%
$^{2}$Carnegie Mellon University\qquad%
$^{3}$IBM Research%
}\vspace{2pt}\\{\it%
$^{4}$Georgia Institute of Technology\qquad
$^{5}$NVIDIA Research\qquad
$^{6}$ETH Z{\"u}rich}
}

\begin{document}
\maketitle

{
\begin{abstract}
% !TEX root = ../paper.tex

\new{This paper summarizes the idea of Low-Cost Interlinked Subarrays (LISA),
  which was published in HPCA 2016~\cite{chang-hpca2016}, and examines
the work's significance and future potential}. \new{Our HPCA 2016}
paper introduces a new DRAM design that enables fast and energy-efficient bulk
data movement across subarrays in a DRAM chip.  While bulk data movement is a
key operation in many applications and operating systems, \new{we observe that}
contemporary systems perform this movement inefficiently, by transferring data
from DRAM to the processor, and then back to DRAM, across a narrow off-chip
channel. The use of this narrow channel for bulk data movement results in high
latency and energy consumption. Prior work proposes to avoid these high costs
by exploiting the \emph{existing} wide internal DRAM bandwidth for bulk data
movement, but the limited connectivity of wires within DRAM allows fast data
movement within only a single DRAM subarray. Each subarray is only a few
megabytes in size, greatly restricting the range over which fast bulk data
movement can happen within DRAM.% a DRAM chip.

\new{Our HPCA 2016 paper} proposes a new DRAM substrate, \capfullname (\lisa),
whose goal is to enable fast and efficient data movement across a large range of
memory at low cost. \lisa adds low-cost connections \emph{between} adjacent
subarrays.  By using these connections to interconnect the existing internal
wires (\emph{bitlines}) of adjacent subarrays, \lisa enables wide-bandwidth data
transfer across multiple subarrays with little (only 0.8\%) DRAM area overhead.
As a DRAM substrate, \lisa is versatile, enabling a \new{variety} of new
applications.  We describe and evaluate three such applications in detail:
(1)~fast inter-subarray bulk data copy, (2)~in-DRAM caching using a DRAM
architecture whose rows have heterogeneous access latencies, and (3)~accelerated
bitline precharging by linking multiple precharge units together. Our extensive
evaluations show that each of LISA's three applications significantly improves
performance and memory energy efficiency, and their combined benefit is higher
than the benefit of each alone, on a variety of workloads and system
configurations.

\end{abstract}

% !TEX root=../paper.tex

\section{Introduction}

Bulk data movement, the movement of thousands or millions of bytes between two
memory locations, is a common operation performed by an increasing number of
real-world applications \newI{(e.g., \cite{kanev-isca2015, lee-hpca2013,
  ousterhout-usenix1990, rosenblum-sosp1995, seshadri-cal2015,
  seshadri-micro2013, son-isca2013, sudan-asplos2010,zhao-iccd2005,
  seshadri-micro2017, lee-taco2016, ami-asplos2018})}. Therefore, it has been the target of
several architectural optimizations \newI{(e.g., \cite{blagodurov-usenix2011,
  jiang-pact2009, seshadri-micro2013, wong-fpt2006, zhao-iccd2005,
  lee-taco2016,seshadri-isca2015,mutlu-imw2013,kim-apbc2018,ami-asplos2018})}. In fact, bulk data movement is
important enough that modern commercial processors are adding specialized
support to improve its performance, such as the ERMSB instruction recently added
to the x86 ISA~\cite{intel-optmanual2012}.

In today's systems, to perform a bulk data movement between two locations in
memory, the data needs to go through the processor \emph{even though both the
source and destination are within memory}. To perform the movement, the data is
first read out one cache line at a time from the source location in
memory into the processor caches, over a pin-limited off-chip channel (typically
64~bits wide). Then, the data is written back to memory, again one cache line at
a time over the pin-limited channel, into the destination location. By going
through the processor, this data movement incurs a significant penalty in terms
of latency and energy consumption.

To address the inefficiencies of traversing the pin-limited channel, a number of
mechanisms have been proposed to accelerate bulk data movement
(e.g.,~\cite{jiang-pact2009,lu-micro2015,seshadri-micro2013,zhao-iccd2005}). The
state-of-the-art mechanism, \rc~\cite{seshadri-micro2013}, performs data
movement \emph{completely within a DRAM chip}, avoiding costly data transfers
over the pin-limited memory channel. However, its effectiveness is limited
because \rc can enable \emph{fast} data movement \emph{only} when the source and
destination are within the same DRAM \emph{subarray}. A DRAM chip is divided
into multiple \emph{banks} (typically 8), each of which is further split into
many \emph{subarrays} (16 to 64)~\cite{kim-isca2012}, shown in
\figref{intro_baseline}, to ensure reasonable read and write latencies at high
density~\cite{ chang-hpca2014,jedec-ddr3, jedec-ddr4, kim-isca2012,
  udipi-isca2010}.\footnote{We refer the reader to our prior
  works\new{~\cite{chang-sigmetrics2016, kim-isca2012, lee-hpca2013,
      lee-hpca2015, kim-micro2010, kim-hpca2010,chang-hpca2016, hassan-hpca2016,
      chang-sigmetrics2017, lee-sigmetrics2017, lee-taco2016, lee-pact2015,
      liu-isca2012, liu-isca2013, patel-isca2017, chang-hpca2014,
      seshadri-micro2013, seshadri-micro2017, hassan-hpca2017, kim-cal2015,
      kim-isca2014, kim-hpca2018}} for a detailed background on
  DRAM.} Each subarray is a two-dimensional array with hundreds of rows of DRAM
cells, and contains only a few megabytes of data (e.g., 4MB in a rank of eight
1Gb~DDR3 DRAM chips with 32 subarrays per bank). While two DRAM rows in the
\emph{same} subarray are connected via a wide (e.g., 8K~bits) bitline interface,
rows in \emph{different} subarrays are connected via only a \emph{narrow 64-bit
  data bus} within the DRAM chip (\figref{intro_baseline}). Therefore, even for
previously-proposed in-DRAM data movement mechanisms such as
\rc~\cite{seshadri-micro2013}, \emph{inter-subarray} bulk data movement incurs
long latency and high memory energy consumption even though data does \emph{not}
move out of the DRAM chip.

%\captionsetup[subfloat]{font={small},margin={-0.4in,0in}} % Old template
\begin{figure}[h]
    \vspace{-5pt}
    \centering
    \subfloat[RowClone~\cite{seshadri-micro2013}]{
        \begin{minipage}[t][2.5cm]{\linewidth/2-0.08in}
        \begin{center}
        \label{fig:intro_baseline}
        \includegraphics[scale=0.92]{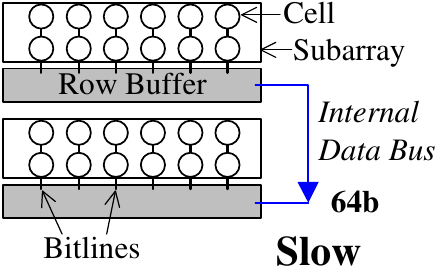}
        \end{center}
        \end{minipage}
    }
    \hfill
    \subfloat[\acro]{
        \begin{minipage}[t][2.5cm]{\linewidth/2-0.22in}
        \begin{center}
        \label{fig:intro_lisa}
        \includegraphics[scale=0.92]{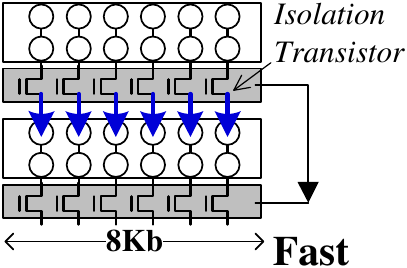}
        \end{center}
        \end{minipage}
    }%

    \caption{Transferring data between subarrays using the internal data bus
        takes a long time in state-of-the-art DRAM design,
        \rc~\cite{seshadri-micro2013} (a). Our work, LISA, enables fast
        inter-subarray data movement with a low-cost substrate (b). Reproduced
        from~\cite{chang-hpca2016}.}

    \label{fig:intro}
\end{figure}
\captionsetup[subfloat]{font={small},margin={0in,0in}}
 %-- 2 columns
%\figputHS{intro_slow_camera}{0.92}{Transferring data between subarrays using the
%  internal data bus takes a long time in state-of-the-art DRAM design,
%  \rc~\cite{seshadri-micro2013}. Reproduced from~\cite{chang-hpca2016}.}

While it is clear that fast \emph{inter-subarray} data movement can have several
applications that improve system performance and memory energy
efficiency\newI{~\cite{kanev-isca2015, ousterhout-usenix1990,
  rosenblum-sosp1995,seshadri-cal2015,seshadri-micro2013,zhao-iccd2005,seshadri-micro2017,
  lee-pact2015,ami-asplos2018}}, there is currently no mechanism that performs such data
movement quickly and efficiently.  This is because \emph{no wide datapath exists
  today between subarrays} within the same bank (i.e., the connectivity of
subarrays is low in modern DRAM). \textbf{Our goal} is to design a low-cost DRAM
substrate that enables fast and energy-efficient data movement \emph{across
  subarrays}.

% !TEX root=../paper.tex

\section{Low-Cost Inter-Linked Subarrays (LISA)}
\label{sec:lisa}

We make two key observations that allow us to improve the connectivity of
subarrays within each bank in modern DRAM. First, accessing data in DRAM causes
the transfer of an entire row of DRAM cells to a buffer (i.e., the \emph{row
  buffer}, where the row data temporarily resides while it is read or written)
via the subarray's \emph{bitlines}. Each bitline connects a column of cells to
the row buffer, interconnecting every row within the same subarray
(\figref{intro_baseline}). Therefore, the bitlines essentially serve as a very
wide bus that transfers a \emph{row's worth of data} (e.g., 8K~bits in a chip)
at once. Second, subarrays within the same bank are placed in close proximity to
each other. Thus, the bitlines of a subarray are very close to (but are not
currently connected to) the bitlines of neighboring subarrays (as shown in
\figref{intro_baseline}).

\textbf{Key Idea.} Based on these two observations, we introduce a new
DRAM substrate, called \emph{\lisafullname} (\emph{\lisa}). \lisa enables
\emph{low-latency, high-bandwidth inter-subarray connectivity} by linking
neighboring subarrays' bitlines together with \emph{isolation transistors}, as
illustrated in \figref{intro_lisa}. We use the new inter-subarray connection in
\lisa to develop a new DRAM operation, \emph{\xferfull (\xfer)}, which moves
data that is latched in an activated row buffer in one subarray into an inactive
row buffer in another subarray, without having to send data through the narrow
internal data bus in DRAM. \xfer exploits the fact that
the activated row buffer has enough drive strength to induce charge perturbation
within the idle (i.e., \emph{precharged}) bitlines of neighboring subarrays,
allowing the destination row buffer to sense and latch this data when the
isolation transistors are enabled. We describe the detailed operation of
RBM in our HPCA 2016 paper~\cite{chang-hpca2016}.

%\figputHS{intro_fast_camera}{0.92}{ Our work, LISA, enables fast inter-subarray
%data movement with a low-cost substrate. Reproduced from~\cite{chang-hpca2016}.}

By using a rigorous DRAM circuit model that conforms to the JEDEC
standards~\cite{jedec-ddr3} and ITRS
specifications~\cite{itrs-fep-2013,itrs-interconnect-2013}, we show that \xfer
performs \emph{row buffer movement} at 26x the bandwidth of a modern 64-bit
DDR4-2400 memory channel (500~GB/s vs.\ 19.2~GB/s), even after we conservatively
add a large (60\%) timing margin to account for process and temperature
variation.

%\subsection{Die Area Overhead}
\label{sec:hwarea}

\textbf{Die Area Overhead.} To evaluate the area overhead of adding isolation
transistors, we use area values from prior work, which adds isolation
transistors to disconnect bitlines from sense amplifiers~\cite{o-isca2014}. That
work shows that adding an isolation transistor to every bitline incurs a total
of 0.8\% die area overhead in a 28nm DRAM process technology. Similar to prior
work that adds isolation transistors to DRAM~\cite{lee-hpca2013,o-isca2014}, our
\lisa substrate also requires additional control logic outside the DRAM banks to
control the isolation transistors, which incurs a small amount of area and is
non-intrusive to the cell arrays.

% !TEX root=../paper.tex

\section{Applications of LISA}

We exploit LISA's \emph{fast inter-subarray movement capability} to enable many
applications that can improve system performance and energy efficiency. In
\new{our HPCA 2016 paper~\cite{chang-hpca2016}}, we implement and evaluate three applications of \lisa, which significantly
improve system performance in different ways.

\subsection{Rapid Inter-Subarray Bulk Data Copying (\lisarc)}
\label{ssec:risc}

Due to the narrow memory channel width, bulk copy operations used by
applications and operating systems are performance limiters in today's
systems~\cite{jiang-pact2009, kanev-isca2015, seshadri-micro2013,zhao-iccd2005,
  lee-pact2015}. These operations are commonly performed due to the
\texttt{memcpy} and \texttt{memmov}. Recent work reported that these two
operations consume 4-5\% of \emph{all of} Google's data center
cycles~\cite{kanev-isca2015}, making them an important target for lightweight
hardware acceleration.

Our goal is to design a new mechanism that enables \emph{low-latency} and
\emph{energy-efficient} memory copy between rows \emph{in different subarrays}
within the same bank. To this end, we propose a new in-DRAM copy mechanism that
uses \lisa to exploit the high-bandwidth links between subarrays. The key idea,
step by step, is to: (1)~activate a source row in a subarray; (2)~rapidly
transfer the data in the activated source row buffers to the destination
subarray's row buffers, through LISA's \xfer operation;
%wide inter-subarray links, without using the narrow internal data bus;
and (3)~activate the destination row, which
enables the contents of the destination row buffers to be latched into the
destination row. We call this inter-subarray row-to-row copy mechanism
\emph{\lisa-\underline{R}apid \underline{I}nter-\underline{S}ubarray
  \underline{C}opy} (\lisarc).

%Our detailed SPICE DRAM circuit model shows that \lisarc reduces the copy
%latency by 9x \new{and copy energy by 48x} compared to state-of-the-art work,
%RowClone~\cite{seshadri-micro2013}, when copying 8KB of data between two
%different subarrays within the same bank.

\subsubsection{\new{DRAM Latency and Energy Consumption}}

\new{
\figref{copy_latency_energy} shows the DRAM latency and DRAM energy consumption
of \baseline (i.e, the baseline system), RowClone~\cite{seshadri-micro2013}
(state-of-the-art work), and \lisarc for copying a row of data (8KB). The exact
latency and energy numbers are listed in \tabref{copy_latency}. For \lisarc, we
define a \emph{hop} as the number of subarrays that \lisarc needs to copy data
\emph{across} to move the data from the source subarray to the destination
subarray. For example, if the source and destination subarrays are adjacent to
each other, the number of hops is 1. The DRAM chips \newI{we} evaluate have 16
subarrays per bank, so the maximum number of hops is 15.

\figputGTW{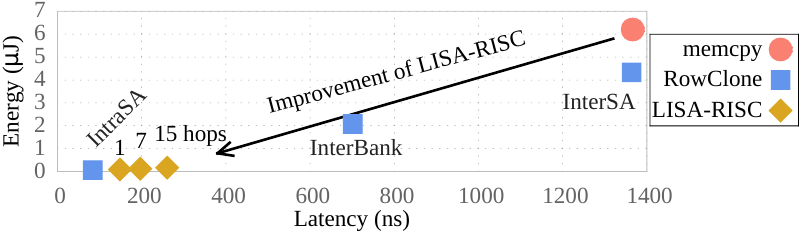}{Latency and DRAM energy of 8KB copy.
  Reproduced from \cite{chang-hpca2016}.}
%\begin{table}[t]
%    \setlength{\tabcolsep}{.55em}
%    \centering
%
%    \begin{tabular}{lrr}
%        \toprule
%        \textbf{Copy Mechanisms} & \textbf{Latency} (ns) & \textbf{Energy} ($\micro$J)\\
%        \cmidrule(rl){1-3}
%
%        \baseline (via mem. channel) & 1366 & 6.20 \\
%        RowClone~\cite{seshadri-micro2013} & 1364 & 4.33 \\
%        \lisarc & 149 & 0.09 \\
%        \bottomrule
%    \end{tabular}
%
%    \caption{Latency and DRAM energy due to coping 8KB data across two different
%subarrays within the same bank.}
%
%    \label{tab:copy_latency}%
%\end{table}

\begin{table}[t]
    % Table specifics -- font size and spacing
    \fontsize{8.5}{10.2}\selectfont
    \newdimen\origwspace
    \origwspace=\fontdimen2\font% original inter-word space
    \fontdimen2\font=0.22em% decrease inter word space
    \setlength{\tabcolsep}{.2em}

    \centering
    \begin{tabular}{lll}
        \toprule
        \textbf{Copy Commands} (8KB) & \textbf{Latency} (ns) & \textbf{Energy} ($\micro$J)\\
        \cmidrule(rl){1-3}
        \baseline (via mem. channel) & 1366.25 & 6.2 \\
        RC-InterSA / Bank / IntraSA & 1363.75 / 701.25 / 83.75 & 4.33 / 2.08 / 0.06\\
        \lisarc (1 / 7 / 15 hops) & 148.5 / 196.5 / 260.5 & 0.09 / 0.12 / 0.17 \\
        \bottomrule
    \end{tabular}
    \caption{Copy latency and DRAM energy. \new{Reproduced from
        \cite{chang-hpca2016}}.}
    \label{tab:copy_latency}%
    \fontdimen2\font=\origwspace
\end{table}

We make two observations from these numbers. First, although inter-subarray
RowClone (\emph{\rcirsa}) incurs similar latencies as \baseline, it consumes
1.43x less energy, as it does \emph{not} transfer data over the channel and DRAM
I/O for each copy operation. However, as we discuss in Section 4.1 of our
HPCA 2016 paper~\cite{chang-hpca2016}, \rcirsa incurs a higher system
performance penalty because it is a \newI{\emph{blocking}} long-latency  memory command.
Second, copying between subarrays using \lisa reduces the copy latency by 9x and
copy energy by 48x compared to \rc, even though the total latency of \lisarc
grows linearly with the hop count. An additional benefit of using \lisarc is
that its inter-subarray copy operations are performed \emph{completely inside a
  bank}. As the internal DRAM data bus is untouched, \emph{other} banks can
\newI{\emph{concurrently}} serve memory requests, exploiting bank-level parallelism. }

%By exploiting the \lisa substrate, we thus provide a more complete set of
%in-DRAM copy mechanisms.

%\tabref{copy_latency} summarizes the
%exact latency and DRAM energy needed for copying 8KB of data in today's baseline
%system~\cite{chang-hpca2016}, RowClone~\cite{seshadri-micro2013}, and \lisarc.

\subsubsection{Evaluation}

We briefly summarize the system performance improvement due to \lisarc on a
quad-core system. We evaluate our system using
Ramulator~\cite{kim-cal2015,ram-github}, an open-source cycle-accurate DRAM
simulator, driven by traces generated from Pin~\cite{luk-pldi2005}. Our workload
evaluation results show that \lisarc outperforms RowClone and \baseline: its
average performance improvement and energy reduction over the best performing
inter-subarray copy mechanism (i.e., \baseline) are 66.2\% and 55.4\%,
respectively, on a quad-core system, across 50 workloads that perform bulk
copies. \new{We refer the reader to Section 9 of our HPCA 2016 paper}~\cite{chang-hpca2016}
for \newI{detailed evaluation and analysis.}

\subsection{In-DRAM Caching Using \\ Heterogeneous Subarrays (\lisavilla)}

Our second application aims to reduce the DRAM access latency for
frequently-accessed (hot) data. We propose to introduce heterogeneity
\emph{within a bank} by designing \emph{heterogeneous-latency subarrays}. We
call this heterogeneous DRAM design
\emph{\underline{V}ar\underline{I}ab\underline{L}e \underline{LA}tency DRAM}
(\villa). To design a low-cost fast subarray, we take an approach similar to
prior work, attaching fewer cells to each bitline to reduce the parasitic
capacitance and resistance. This reduces the latency of the three fundamental
DRAM operations--\emph{activation}, \emph{precharge}, and
\emph{restoration}--when accessing data in the fast
subarrays~\cite{lee-hpca2013, micron-rldram3, son-isca2013}. \emph{Activation}
``opens'' a row of DRAM cells to access stored data. \emph{Precharge} ``closes''
an activated row. \emph{Restoration} restores the charge level of each DRAM cell
in a row to prevent data loss. Together, these three operations predominantly
define the latency of a memory request\newI{~\cite{chang-sigmetrics2016, kim-isca2012, lee-hpca2013,
      lee-hpca2015, kim-micro2010, kim-hpca2010,chang-hpca2016, hassan-hpca2016,
      chang-sigmetrics2017, lee-sigmetrics2017, lee-taco2016, lee-pact2015,
      liu-isca2012, liu-isca2013, patel-isca2017, chang-hpca2014,
      seshadri-micro2013, seshadri-micro2017, hassan-hpca2017, kim-cal2015,
      kim-isca2014, kim-hpca2018}}. In this work, we focus
on managing the fast subarrays in hardware, as \new{doing so} offers better
adaptivity to \newI{\emph{dynamic}} changes in the hot data set.

In order to take advantage of \villa, we rely on \lisarc to rapidly copy rows
across subarrays, which significantly reduces the caching latency. We call this
synergistic design, which builds \villa using \lisa,
\emph{\lisavilla}. Nonetheless, the cost of transferring data to a fast
subarray is still non-negligible, especially if the fast subarray is far from
the subarray where the data to be cached resides. Therefore, an intelligent
cost-aware mechanism is required to make astute decisions on which data to
cache and when.

\subsubsection{Caching Policy for \lisavilla}
\label{sec:villa:caching}

We design a simple epoch-based caching policy to evaluate the benefits of
caching a row in \lisavilla. Every epoch, we track the number of accesses to
rows by using a set of 1024 saturating counters for each bank.\footnote{The
  hardware cost of these counters is low, requiring only 6KB of storage in the
  memory controller (see Section 7.1 of our HPCA 2016
  paper~\cite{chang-hpca2016}).} The counter values are halved every epoch to
prevent staleness. At the end of an epoch, we mark the 16 most
frequently-accessed rows as \emph{hot}, and cache them when they are accessed
the next time. For our cache replacement policy, we use the \emph{benefit-based
  caching} policy proposed by Lee et al.~\cite{lee-hpca2013}. Specifically, it
uses a benefit counter for each row cached in the fast subarray: whenever a
cached row is accessed, its counter is incremented. The row with the least
benefit is replaced when a new row needs to be inserted. Note that a large body
of work \newI{proposes} various caching policies \new{(e.g., \cite{hart-compcon1994,
    hidaka-ieeemicro90,hsu-isca1993,jiang-hpca2010,kedem-1997,meza-cal2012,qureshi-isca2007,
    seshadri-pact2012, yoon-iccd2012, seshadri-taco2015,
    qureshi-isca2006,tyson-micro1995,li-cluster2017, yu-micro2017})}, each of
which can potentially be used with \lisavilla.

\subsubsection{Evaluation}

\figref{villa_multic_errbar} shows the system performance improvement of
\lisavilla over a baseline without any fast subarrays in a four-core system. It
also shows the hit rate in \villa, i.e., the fraction of accesses that hit in
the fast subarrays. We make two main observations. First, by exploiting \lisarc
to quickly cache data in \villa, \lisavilla improves system performance for a
wide variety of workloads --- by up to 16.1\%, with a geometric mean of 5.1\%.
This is mainly due to reduced DRAM latency of accesses that hit in the fast
subarrays. The performance improvement heavily correlates with the VILLA cache
hit rate. Second, the \villa design, which consists of heterogeneous subarrays,
is not practical without \lisa. \figref{villa_multic_errbar} shows that using
\rcirsa (i.e., RowClone copying data across subarrays) to move data into the
cache \emph{reduces} performance by 52.3\% due to slow data movement, which
overshadows the benefits of caching. The results indicate that \lisa is an
important substrate to enable not only fast bulk data copy, but also a fast
in-DRAM caching scheme.

\figputGHWHeight{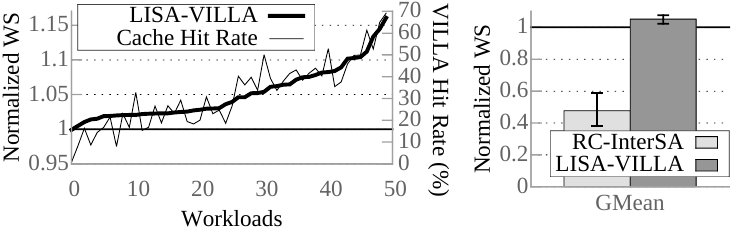}{Performance improvement and hit rate with
\lisavilla, and performance comparison to using \rcirsa with \villa. Reproduced
from~\cite{chang-hpca2016}.}{1}

\subsection{Fast Precharge Using Linked Precharge Units (\lisapre)}
\label{sec:lisapre}

Our third application aims to accelerate the process of precharge. The precharge
time for a subarray is determined by the drive strength of the precharge unit
(i.e., a circuitry in a subarray's row buffer for precharging the connected
subarray). We observe that in modern DRAM, while a subarray is being precharged,
the precharge units (PUs) of \emph{other} subarrays remain idle.

We propose to exploit these idle PUs to accelerate a precharge operation by
connecting them to the subarray that is being precharged. Our mechanism,
\emph{\lisa-\underline{LI}nked \underline{P}recharge} (\lisapre), precharges a
subarray using \emph{two} sets of PUs: one from the row buffer that is being
precharged, and a second set from a neighboring subarray's row buffer (which is
already in the precharged state), by enabling the links between the two
subarrays.

To evaluate the accelerated precharge process, we use the same DRAM circuit
model described in Section~\ref{sec:lisa} and simulate the linked precharge
operation in SPICE. Our SPICE simulation reports that \lisapre significantly
reduces the precharge latency by 2.6x compared to the baseline (5ns vs. 13ns).
Our system evaluation shows that \mbox{\lisapre} improves performance by 10.3\%
on average, across 50 \mbox{four-core} workloads. \new{We refer the reader to
  Section 6 of our HPCA 2016 paper~\cite{chang-hpca2016} for a detailed analysis
  of LISA-LIP.}

\subsection{Evaluation: Putting Everything Together}
\label{sec:everything_results}

As all of the three proposed applications are complementary to each other, we
evaluate the effect of putting them together on a four-core system.
\figref{all_norm_ws_4c} shows the system performance improvement of adding
\lisavilla to \lisarc, as well as combining all three optimizations, compared to
our baseline using \baseline and standard DDR3-1600 memory across 50 workloads.
We refer the reader to our full paper~\cite{chang-hpca2016} for the detailed
configuration and workloads. We draw several key conclusions. First, the
performance benefits from each scheme are additive. On average, adding
\lisavilla improves performance by 16.5\% over \lisarc alone, and adding
\lisapre further provides an 8.8\% gain over \lisarcvilla. Second, although
\lisarc alone provides a majority of the performance improvement over the
baseline (59.6\% on average), the use of both \mbox{\lisavilla} and
\mbox{\lisapre} further improves performance, resulting in an average
performance gain of 94.8\% and memory energy reduction (not plotted) of 49.0\%.
Taken together, these results indicate that \lisa is an effective substrate that
enables a wide range of high-performance and energy-efficient applications in
the DRAM system.

%\figputGHW{all_norm_ws_4c}{Combined WS improvement of \lisa applications.}

\figputGHWHeight{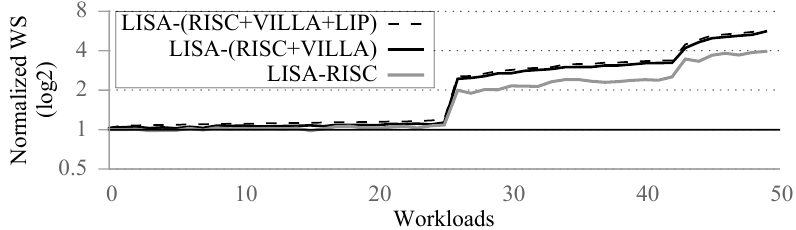}{Combined weighted speedup
  (WS)~\cite{eyerman-ieeemicro2008,snavely-asplos2000} improvement of \lisa
  applications. Reproduced from~\cite{chang-hpca2016}.}{0.97}

\new{We conclude that LISA is an effective substrate that can greatly improve
  system performance and reduce system energy consumption by synergistically
  enabling multiple different applications. Our HPCA 2016 paper~\cite{chang-hpca2016} provides many
  more experimental results and analyses confirming this finding.}

% !TEX root = ../paper.tex

\section{Related Work}
\label{sec:related}

To our knowledge, this is the first work to propose a DRAM substrate that
supports fast data movement between subarrays in the same bank, which enables a
wide variety of applications for DRAM systems. We now discuss prior works that
focus on each of the optimizations that \lisa enables.

%\vspace{3pt} \noindent\textbf{Bulk Data Transfer Mechanisms.}
\subsection{Bulk Data Transfer Mechanisms}

Prior works~\cite{gschwind-cf2006, gummaraju-pact2007,
kahle-ibmjrd2005,carter-hpca1999,zhang-ieee2001} propose to add scratchpad
memories to reduce CPU pressure during bulk data transfers, which can also
enable sophisticated data movement (e.g.,
scatter-gather~\cite{seshadri-micro2015}), but they still require data to first
be moved on-chip. A patent proposes a DRAM \new{design} that can copy a page
across memory blocks~\cite{seo-patent}, but lacks concrete analysis and
evaluation of the underlying copy operations. Intel I/O Acceleration
Technology~\cite{intelioat} allows for memory-to-memory DMA transfers
\emph{across a network}, but cannot transfer data within main memory.

Zhao et al.~\cite{zhao-iccd2005} propose to add a bulk data movement engine inside the memory
controller to speed up bulk-copy operations. Jiang et
al.~\cite{jiang-pact2009} design a different copy engine, placed within the
cache controller, to alleviate pipeline and cache stalls that occur when these
transfers take place. However, these works do not directly address the problem
of data movement \new{across} the narrow memory channel.

A concurrent work by Lu et al.~\cite{lu-micro2015} proposes a heterogeneous DRAM
design similar to \villa, called \mbox{DAS-DRAM}, but with a very different data
movement mechanism from LISA. It introduces a row of \emph{migration cells} into
each subarray to move rows across subarrays. Unfortunately, the latency of
DAS-DRAM is not scalable with movement distance, because it requires writing the
migrating row into each intermediate subarray's migration cells before the row
reaches its destination, which prolongs data transfer latency. In contrast, LISA
provides a \emph{direct path} to transfer data \emph{between row buffers}
\new{between adjacent subarrays} without requiring intermediate data writes into
any subarray.

\vspace{3pt}

%\noindent\textbf{Cached DRAM.}
\subsection{Cached DRAM}

Several prior works (e.g., \cite{ hart-compcon1994, hidaka-ieeemicro90,
hsu-isca1993,kedem-1997,zhang-ieeemicro2001}) propose to add a small SRAM cache
to a DRAM chip to lower the access latency for data that is kept in the SRAM
cache (e.g., frequently or recently used data). There are two main disadvantages
of these works. First, adding an SRAM cache into a DRAM chip is very intrusive:
it incurs a high area overhead (38.8\% for 64KB in a 2Gb DRAM chip) and design
complexity~\cite{lee-hpca2013,kim-isca2012}. Second, transferring data from DRAM
to SRAM uses a narrow global data bus, internal to the DRAM chip, which is
typically 64-bit wide. Thus, installing data into the DRAM cache incurs high
latency. Compared to these works, our LISA-VILLA design enables low latency
without significant area overhead or complexity.

\subsection{Heterogeneous-Latency DRAM}

Prior works propose DRAM architectures that provide heterogeneous latency
either \emph{spatially} (dependent on \emph{where} in the memory an access
targets) or \emph{temporally} (dependent on \emph{when} an access occurs).

\textbf{Spatial Heterogeneity.}
Prior work introduces spatial heterogeneity into DRAM, where one region has a
fast access latency but fewer DRAM rows, while the other has a slower access
latency but many more rows~\cite{lee-hpca2013, son-isca2013}. \new{Recent works
  show that latency heterogeneity inherent in DRAM chips due to process or
  design-induced variation can also naturally enable such heterogeneous-latency
  substrates~\cite{chang-sigmetrics2016,lee-sigmetrics2017}.} The fast region
\new{in DRAM can be} utilized as a caching area, for the frequently or recently
accessed data. We briefly describe two state-of-the-art works that offer
different heterogeneous-latency DRAM designs.

CHARM~\cite{son-isca2013} introduces heterogeneity \emph{within a rank} by
designing a few fast banks with (1)~shorter bitlines for faster data sensing,
and (2)~closer placement to the chip I/O for faster data transfers. To exploit
these low-latency banks, CHARM uses an OS-managed mechanism to \emph{statically}
map hot data to these banks, based on profiled information from the compiler or
programmers. Unfortunately, this approach \emph{cannot adapt} to program phase
changes, limiting its performance gains. If it were to adopt dynamic hot data
management, CHARM would incur high migration costs over the narrow 64-bit bus
that internally connects the fast and slow banks.

TL-DRAM~\cite{lee-hpca2013} provides heterogeneity \emph{within a subarray} by
dividing it into fast (near) and slow (far) segments that have short and long
bitlines, respectively, using isolation transistors. The fast segment can be
managed as an OS-transparent hardware cache. The main disadvantage is that it
needs to cache each hot row in \emph{two near segments} as each subarray uses
two row buffers on \emph{opposite ends} to sense data in the open-bitline
architecture (as discussed in our HPCA 2016 paper~\cite{chang-hpca2016}). This
prevents TL-DRAM from using the full near segment capacity. As we can see,
neither CHARM nor TL-DRAM strike a good design balance for heterogeneous-latency
DRAM. Our proposal, LISA-VILLA, is a new heterogeneous DRAM design that offers
fast data movement with a low-cost and easy-to-implement design.

\textbf{Temporal Heterogeneity.}
Prior work observes that DRAM latency can vary depending on \emph{when} an
access occurs. The key observation is that a \newI{\emph{recently-accessed or
  refreshed}} row has nearly full electrical charge in the cells, and thus the
following access to the same row can be performed faster~\cite{hassan-hpca2016,
  hassan-hpca2017, shin-hpca2014}. We briefly describe two state-of-the-art
works that focus on providing heterogeneous latency temporally.

ChargeCache~\cite{hassan-hpca2016} enables faster access to
\emph{recently-accessed} rows in DRAM by tracking the addresses of
recently-accessed rows. NUAT~\cite{shin-hpca2014} enables accesses to
recently-refreshed rows at low latency because these rows are already
highly-charged. 
% The main issue with these works is that the proposed effect of
% highly-charged cells can be accessed with lower latency, is slightly observable
% only when very long refresh intervals are used on \emph{existing} DRAM chips, as
% demonstrated by a recent characterization work~\cite{hassan-hpca2017}. However,
% within the duration of the standard 64ms refresh interval, no latency benefits
% can be directly observed on \emph{existing} DRAM chips. 
\newI{In contrast to ChargeCache and NUAT, LISA does not require
data to be recently-accessed/refreshed in order to reduce DRAM latency.}
\new{Adaptive-Latency DRAM (AL-DRAM)~\cite{lee-hpca2015} adapts the DRAM latency of
  each DRAM module to temperature, observing that each module can be operated
  faster at lower temperatures. LISA is orthogonal to AL-DRAM. \newI{The
  ideas} of LISA can be employed in conjunction with works that exploit the temporal
  heterogeneity of DRAM latency.}

\subsection{Other Latency Reduction Mechanisms}

\newI{Many prior works propose memory scheduling techniques, which generally
reduce latency to access DRAM~\cite{kim-hpca2010, kim-micro2010,
  subramanian-tpds2016, subramanian-iccd2014, mutlu-isca2008, mutlu-micro2007,
  usui-taco2016, ausavarungnirun-isca2012, moscibroda-usenix2007,ebrahimi-micro2011,subramanian-hpca2013,lee-micro2008,lee-micro2009,ipek-isca2008,morse-hpca12,ghose-isca2013,lee-tech2010}. Other works
propose mechanisms to perform in-memory computation to reduce data movement and
access latency\newI{~\cite{hsieh-iccd2016, seshadri-micro2017, amirali-cal2016, hsieh-isca2016,kim-apbc2018,ami-asplos2018,seshadri-micro2013,
4115697, 592312, guo-wondp14, ahn-isca2015-2, ahn-isca2015, zhang-2014, pattnaik-pact2016, liu-spaa2017, stone-1970}}. LISA is complementary to these works, and it can work
synergistically with in-memory computation mechanisms by enabling fast
aggregation of data.}
\section{Significance}

\new{Our HPCA 2016 paper~\cite{chang-hpca2016} proposes a new DRAM substrate
that significantly improves the performance and efficiency of bulk data 
movement in modern systems.  In this section, we briefly discuss the expected
future impact of our work, and discuss several research directions that our
work motivates.}

\subsection{Potential Industry Impact}

We believe that our LISA substrate can have a large impact on \new{mobile
  systems as well as} data centers that consume a significant amount of cycle
time performing bulk data movement. A recent study~\cite{kanev-isca2015} by
Google reports that \texttt{memcpy()} and \texttt{memmove()} library functions
alone represent 4-5\% of their data center cycles even though Google has a
significant workload diversity running within their data centers. \new{Another
  recent study shows that 62.7\% of system energy is spent on data movement on 
\newI{consumer devices (e.g., smartphones, wearable devices, web-based computers
such as Chromebooks)}~\cite{ami-asplos2018}.} In this work, we demonstrate that one
potential application of using the LISA substrate is to accelerate
\texttt{memcpy()} and \texttt{memmove()}, as discussed in \ssecref{risc}. Our
detailed DRAM circuit model reports that LISA reduces the latency and DRAM
energy of these functions by 9x and 69x compared to today's systems,
respectively. \new{Hence, we expect LISA can improve the efficiency and
  performance of \newI{\emph{both}} mobile and data center systems.}

\subsection{Future Research Directions}

This work opens up several avenues of future research directions. In this
section, we describe several directions that can \new{enable researchers to}
tackle other problems related to memory systems based on the LISA substrate.

\textbf{Reducing Subarray Conflicts via Remapping.} When
two memory requests access two different rows in the same bank, they have to be
served serially, even if they are to different subarrays. To mitigate such
\emph{bank conflicts}, Kim et al.~\cite{kim-isca2012} propose
\emph{subarray-level parallelism (SALP)}, which enables multiple subarrays to
remain activated at the same time. However, if two accesses are to the same
subarray, they still have to be served serially. This problem is exacerbated
when \mbox{frequently-accessed} rows reside in the same subarray. To help
alleviate such \emph{subarray conflicts}, \lisa can enable a simple mechanism
that efficiently remaps or moves the conflicting rows to different subarrays by
exploiting fast \xfer operations.

\textbf{Enabling LISA to Perform 1-to-N Memory Copy or Move Operations.}
A typical \texttt{memcpy} or \texttt{memmove} call only allows the data to be
copied from one source location to one destination location. To copy or move
data from one source location to multiple different destinations, repeated calls
are required. The problem is that such repeated calls incur long latency and
high bandwidth consumption. One potential application that can be enabled by
LISA is performing \texttt{memcpy} or \texttt{memmove} from one source location
to \emph{multiple destinations} completely in DRAM without requiring multiple
calls of these operations.

By using LISA, we observe that moving data from the source subarray to the
destination subarray latches the source row's data in all the intermediate
subarrays' row buffer. As a result, activating these intermediate subarrays
would copy their row buffers' data into the specified row within these
subarrays. By extending LISA to perform multi-point (1-to-N) copy or move
operations, we can significantly increase system performance of several
commonly-used system operations. For example, forking multiple child processes
can utilize 1-to-N copy operations to efficiently copy memory pages from the
parent's address space to all the children. \newI{As another example, LISA can extend} \new{the range of in-DRAM
bulk bitwise operations\newI{~\cite{seshadri-micro2017,seshadri-cal2015}}}. \new{Thus, LISA can
efficiently enable architectural support to a new, useful system and
programming primitive: 1-to-N bulk memory copy/movement.}

\textbf{In-Memory Computation with LISA.}
One important requirement of efficient in-memory computation is being able to
move data from its stored location to the computation units with very low
latency and energy. We believe using the LISA substrate can enable a new
in-memory computation framework. The idea is to add a small computation unit
inside each or a subset of banks, and connect these computation units to the
neighboring subarrays which store the data. Doing so allows the system to
utilize LISA to move bulk data from the subarrays to the computation units
\new{with low} latency and low area overhead.

%Two potential types of computation units to add are bitwise shifters and
%ripple-carry adders since simple integer addition and bitwise shifting between
%two arrays of data are common operations in many applications. One key challenge
%of adding computation units would be fitting each single unit that processes a
%single bit within a pitch of DRAM array's column. For example, a single-bit
%shifter requires 12 transistors which is much bigger than a sense amplifier (4
%transistors). This implementation overhead can restrict the computation to
%process data at the granularity of a row size. Nonetheless, this general
%in-memory computation framework still has the potential to enable simple
%filtering operations in memory to provide high system performance or energy
%efficiency at low cost.

\textbf{Extending LISA to Non-Volatile Memory.}
In this work, we only focus on the DRAM technology. A class of emerging memory
technology is non-volatile memory (NVM), which has the capability of retaining
data without power supply. We believe that the LISA substrate can be extended to
NVM \new{(e.g., PCM~\cite{lee-isca2009, lee-ieeemicro2010, qureshi-isca2009,
    yoon-taco2014,lee-cacm2010,qureshi-micro2009,yoon-iccd2012} and
  \mbox{STT-MRAM}~\cite{ku-ispass2013,guo-isca2009,chang-hpca2013})} since the memory
organization of NVM mostly resembles that of DRAM. A potential application of
LISA in NVM is an efficient file copy operation that does not incur costly I/O
data transfer. \new{We believe LISA can provide further benefits when main
  memory becomes persistent~\cite{meza-weed2013}.}

% !TEX root = ../paper.tex

\section{Conclusion} \label{sec:conclusion}

We present a new DRAM substrate, \emph{\fullname (\lisa)}, that expedites bulk
data movement across subarrays in DRAM. \lisa achieves this by creating a new
high-bandwidth datapath at low cost between subarrays, via the insertion of a
small number of isolation transistors. We describe and evaluate three
applications that are enabled by \lisa. First, \lisa significantly reduces the
latency and memory energy consumption of bulk copy operations between subarrays
over state-of-the-art mechanisms~\cite{seshadri-micro2013}. Second, \lisa
enables an effective in-DRAM caching scheme on a new heterogeneous DRAM
organization, which uses fast subarrays for caching hot data in every bank.
Third, we reduce precharge latency by connecting two precharge units of
adjacent subarrays together using \lisa. We experimentally show that the three
applications of LISA greatly improve system performance and memory energy
efficiency when used individually or together, across a variety of workloads
and system configurations.

We conclude that \lisa is an effective substrate that enables several effective
applications. We believe that this substrate, which enables low-cost
interconnections between DRAM subarrays, can pave the way for other applications
that can further improve system performance and energy efficiency through fast
data movement in DRAM. \new{We greatly encourage future work to 1) investigate
  new applications and benefits of LISA, and 2) develop new low-cost
  interconnection substrates within a DRAM chip to improve \newI{internal connectivity and} data transfer
  ability.}

\section*{Acknowledgments}
We thank the anonymous reviewers and SAFARI group members for their helpful
feedback. We acknowledge the support of Google, Intel, NVIDIA, Samsung, and
VMware. This research was supported in part by the ISTC-CC, SRC, CFAR, and NSF
(grants 1212962, 1319587, and 1320531). Kevin Chang was supported in part by the
SRCEA/Intel Fellowship.

}

\balance
\Urlmuskip=0mu plus 1mu\relax
{
\bstctlcite{bstctl:etal, bstctl:nodash, bstctl:simpurl}
\bibliographystyle{IEEEtranS}
\bibliography{kevin_paper}
}

\end{document}